\documentstyle[aps,amssymb,preprint]{revtex}
 
\begin{document}
\title{Broadcasting of entanglement in three-particle GHZ state \\via quantum copying}
\author{Zhao-Yang Tong and  Le-Man Kuang$^{\dagger}$} 
\address{ Department of Physics, Hunan Normal University, Changsha 410081, China}
\maketitle
\begin{abstract}
 
  We introduce  entanglement measures to describe entanglement in a three-particle system and apply it 
  to studying broadcasting of entanglement in three-particle GHZ state. We show that entanglement of 
  three-qubit GHZ state can be partially broadcasted  with the help of local or non-local copying 
  processes.  It is found that non-local cloning is much more
   efficient than local cloning for the broadcasting of entanglement. 

\noindent PACS number(s): 03.75.Fi, 03.65Bz; 32.80.Pj, 74.20.De
\end{abstract}



   Quantum entanglement, first noted by Einstein-Podolsky-Rosen  and Schr\"{o}dinger [1], is one of the essential 
features of quantum  mechanics.   It has been  well known that  quntum entanglement    plays a key  role 
in many such applications like  quantum teleportation [2], super-dense coding [3], quantum error correction [4], and 
quantum computational speedups [5].  Recently, maximally  entangled states 
of three particles, i.e.,  three-particle GHZ states,  have been produced experimentally [6]. Usually, it is difficult 
to obtain ideal entangled multi-particle state.  An interesting question is: whether there are ways which can broadcast 
the entanglement of correlated systems?  The answer is ok. Masiak and Knight [7] have  shown that  copies of entangled 
pair of qubits can be  genetated through using a universal quantum cloning machine (UQCM) [8] , although  the degree of 
entanglement of the resultant copies is substantially reduced  due to 
a residual entanglement   between the copied output and the copying machine.    On the other hand, up to now there is 
not an appropriate measure to describe quantatively the entanglement of three and more subsystems due to the high 
complexity of entanglement  in multi-particle system [9]. 
The purpose of this paper is to  propose  measures of entanglement in three qubit system in terms of the entanglement 
tensor  approach [10], and use  these measures  to investigate broadcasting of entanglement [11] in three-particle GHZ state.


Consider a system consisting of three qubits. The density operator of the  system can be expanded as a sum  of 
 tensor products in terms of Puali matrices,
\begin{eqnarray}
\hat{\rho}&=&\frac {1}{8}[\hat{1}\otimes\hat{1}\otimes\hat{1}
+\sum_{i=1}^3\lambda_i(1)(\hat{\sigma_i}\otimes\hat{1}\otimes\hat{1}) \nonumber  \\
& &+\sum_{j=1}^3\lambda_j(2)(\hat{1}\otimes\hat{\sigma_j}\otimes\hat{1})+\sum_{k=1}^3\lambda
_k(3)(\hat{1}\otimes\hat{1}\otimes\hat{\sigma_k})\nonumber \\
& &+\sum_{i,k=1}^3K_{ik}(1,3)(\hat{\sigma_i}\otimes\hat{1}\otimes\hat{\sigma_k}) +\sum_{i,j=1}^3K_{ij}(1,2)
    \nonumber \\
& &\times (\hat{\sigma_i}\otimes\hat{\sigma_j}\otimes\hat{1}) +\sum_{j,k=1}^3K_{jk}(2,3)(\hat{1}\otimes\hat{\sigma_j}
\otimes\hat{\sigma_k}) \nonumber \\
& &+\sum_{i,j,k=1}^3K_{ijk}(1,2,3)(\hat{\sigma_i}\otimes\hat{\sigma_j}\otimes\hat{\sigma_k})].
\end{eqnarray}
Here $\{\hat{\sigma}_i, i=1,2,3\}$ are the Pauli matrice.
${\lambda}(1)$,${\lambda}(2)$ and ${\lambda}(3)$ are the three
coherence vectors beloning to qubit 1, 2 and 3, respectively, which determine
the properties of the individual particles.
$K_{ij}(1,2)$, $K_{jk}(2,3)$, and $K_{ik}(1,3)$) are the second-rank correlation tensors
which describe the correlation between qubit 1 and 2 (qubit 2 and 3, qubit 1 and 3) resprctively.
 $K_{ijk}(1,2,3)$ are the three-qubit correlation tensor.

Making use of properties of Pauli matrices  $tr\lbrace\hat{\sigma_i}\rbrace=0$, and 
$tr\lbrace\hat{\sigma_i}\hat{\sigma_j}\rbrace=2\delta_{ij}$,
 we get the following relations:
\begin{eqnarray}
\lambda_i(1)&=&tr(\hat{\rho}\cdot\hat{\sigma_i}\otimes\hat{1}\otimes\hat{1}), \\ 
\lambda_j(2)&=&tr(\hat{\rho}\cdot\hat{1}\otimes\hat{\sigma_j}\otimes\hat{1}), \\
\lambda_k(3)&=&tr(\hat{\rho}\cdot\hat{1}\otimes\hat{1}\otimes\hat{\sigma_k}), \\
K_{ij}(1,2)&=&tr(\hat{\rho}\cdot\hat{\sigma_i}\otimes\hat{\sigma_j}\otimes\hat{1}), \\
K_{ik}(1,3)&=&tr(\hat{\rho}\cdot\hat{\sigma_i}\otimes\hat{1}\otimes\hat{\sigma_k}), \\
K_{jk}(2,3)&=&tr(\hat{\rho}\cdot\hat{1}\otimes\hat{\sigma_j}\otimes\hat{\sigma_k}), \\
K_{ijk}(1,2,3)&=&tr(\hat{\rho}\cdot\hat{\sigma_i}\otimes\hat{\sigma_j}\otimes\hat{\sigma_k}).
\end{eqnarray}

After performing the partial trace over the second qubit   and the third qubit, from Eq.(1) 
we obtain the reduced density operator for the first qubit:  
\begin{equation}
\hat{\rho}^{(1)}=tr_{2,3}(\hat{\rho})=\frac{1}{2}(\hat{1}+\sum_i^3\lambda_i(1)
\hat{\sigma}_i).
\end{equation}
Simmilarly, we can calculate the reduced density operators  $\hat{\rho}^{(2)}$, $\hat{\rho}^{(3)}$ for qubit 2 and 3.
 
  Comparing the direct product $\hat{\rho}^{(1)}\otimes\hat{\rho}^{(2)}\otimes\hat{\rho}^{(3)}$ with Eq.(1),
we can identify the difference by tensors $M(m,n) (1\leq m<n\leq 3)$,
$M(1,2,3)$ defined by 
\begin{equation}
M_{ij}(m,n)=K_{ij}(m,n)-\lambda_i(m)\lambda_j(n),
\end{equation}

\begin{eqnarray}
M_{ijk}(1,2,3)&=&K_{ijk}(1,2,3)\!-\!\lambda_i(1)M_{jk}(2,3)\nonumber  \\
& &-\lambda_j(2)M_{ik}(1,3)-\lambda_k(3)M_{ij}(1,2) \nonumber  \\
& &-\lambda_i(1)\lambda_j(2)\lambda_k(3), 
\end{eqnarray}
which indicates that  $M(1,2,3)=0$   for any product state of three qubits. From the above, 
we also find  that a three-qubit entanglement state necessarilly involves   entanglement between any
two qubits. Hence entanglement measures in a three-qubit system should involve both an  inter-three-qubit entanglement
 measure  and an inter-two-qubit  entanglement  measure. 
 
 Based on $M(1,2,3)$, $M(m,n)(1\leq m<n\leq 3)$, we introduce an inter-three-qubit entanglement
 measure $E_3$ and an inter-two-entanglement measure   $E_2$ in the following form,
\begin{eqnarray}
E_3=\frac{1}{4}\sum_{i,j,k=1}^3M_{ijk}(1,2,3)M_{ijk}(1,2,3), \\
E_2(m,n)=\frac{1}{3}\sum_{i,j=1}^3M_{ij}(m,n)M_{ij}(m,n).
\end{eqnarray}

It is easy to check that $E_2$ and $E_3 $ obey all conditions as entanglement measures indicated in Ref.[10].
 These measures are invariant under local unitary transformations of the subsystems and varies between $0$ 
 (product states) and $1$ (maximum entangled states). $E_3$ quantifies the three-qubit entanglement .
The larger $E_3$ is, the stronger the three-qubit entanglement is. And $E_2(m,n)$
quatitfies the entanglement between any two qubits $m,n$ in the three qubit system.
The larger $E_2(m,n)$ is, the stronger 	the entanglement between two qubit m,n is.

As an example of three-qubit entanglement, let us consider the GHZ state 
$\mid\psi\rangle= (\mid 111\rangle+\mid 000\rangle)/\sqrt{2} $.
Making use of Eqs.(2)-(9), from Eqs.(10) and (11) we can get the  nonzero entanglement tensors, 
$M_{xxx}=1$, $M_{xyy}=M_{yxy}=M_{yxx}=-1$, $M_{zz}(1,2)=M_{zz}(2,3)=M_{zz}(1,3)=1$.
All other entanglement tensors vanish as well as the coherence vectors.  Then from Eqs.(12)and (13) we find that
\begin{equation}
 E_3=1,\quad E_2(1,2)=E_2(2,3)=E_2(1,3)=\frac{1}{3} ,
\end{equation}
which indicate that the GHZ  state is the maximally entangled inter-three-qubit state and contains i
nter-two-qubit entanglement, as expected.

 
 In what follows we  investigate broadcasting of entanglement  in a three-particle GHZ state via quantum copying. 
We shallconsider the two cases  of local cloning and non-local cloning, and assume that the three qubits are prepared in the GHZ state:
\begin{equation}
\mid\psi\rangle=\frac{1}{\sqrt{2}}(\mid 111\rangle_{1_02_03_0}+\mid 000
\rangle_{1_02_03_0}).
\end{equation}

Firstly, we  consider the case  of local cloning. In this case, 
the three qubit $1_0$, $2_0$ and $3_0$ is copied by  the UQCM denoted by
 the   local unitary transformations [8]:
\begin{eqnarray}
U_1\mid 0\rangle_{a_0}\mid 0\rangle_{a_1}\mid X\rangle_x&=&\sqrt{\frac{2}{3}}
\mid 00\rangle_{a_0a_1}\mid \uparrow\rangle_x  \nonumber \\
& &+\sqrt{\frac{1}{3}}\mid +\rangle_{a_0a_1}\mid \downarrow\rangle_x, \\
U_1\mid 1\rangle_{a_0}\mid 0\rangle_{a_1}\mid X\rangle_x&=&\sqrt{\frac{2}{3}}
\mid 11\rangle_{a_0a_1}\mid \downarrow\rangle_x \nonumber \\
& &+\sqrt{\frac{1}{3}}\mid +\rangle_{a_0a_1}\mid \uparrow\rangle_x 
\end{eqnarray}
where $\mid\!+\!\rangle_{a_0a_1}\!=\!(\mid\!10\rangle_{a_0a_1}\!+\!\mid\!01\rangle_{a_0a_1})
/\!\sqrt{2}$. The system labelled by $a_0$ is the original (input) qubit,
while the other system $a_1$ represents the target qubit onto which the information
is copied. The states of the copying machine are labelled by $x$. The state space
of the copying  machine is two dimensional and we assume that it is always in the
same state $\mid X\rangle_x$ initially.

   Suppose that each of the three original qubits $1_0$, $2_0$ and $3_0$
is cloned separately by three distant local cloners $X_1$, $X_2$ and $X_3$. The cloner
$X_1$ ($X_2$,$X_3$) generates out of qubit $1_0$ ($2_0$,$3_0$) two $1_0$ and
$1_1$ ($2_0$ and $2_1$, $3_0$ and $3_1$). After cloning, we get the  following total output
state 
\begin{equation}
\mid\psi\rangle_{total}^{(out)}=\frac{1}{\sqrt{2}}\sum_{i=0}^1\prod_{m=1}^3
[U_1(m)\mid i\rangle_{m_0}\mid 0\rangle_{m_1}\mid X_m\rangle_x].
\end{equation}
After performing trace over the three cloners we obtain a six-qubit density
operator $\hat{\rho}_{1_01_12_02_13_03_1}^{(out)}$ which also describes two
nonlocal three-qubit systems $\hat{\rho}_{1_02_03_0}$ and $\hat{\rho}_{1_1
2_13_1}$. The two three-qubit systems are the clones of the original
three-qubit GHZ state in Eq.(15) and they are described by the density
operators:
\begin{eqnarray}
\hat{\rho}_{1_02_03_0}^{(out)}&=&\frac{7}{24}(\mid 111\rangle\langle 111\mid
+\mid 000\rangle\langle 000\mid) \nonumber \\
& &+\frac{7}{54}(\mid 111\rangle\langle 000\mid
+\mid 000\rangle\langle 111\mid)\nonumber \\
& &+\frac{5}{72}(\mid 110\rangle\langle 110\mid+\mid 011\rangle\langle 011\mid \nonumber  \\
& &+\mid 101\rangle\langle 101\mid+\mid 100\rangle\langle 100\mid \nonumber \\
& &+\mid 010\rangle\langle 010\mid+\mid 001\rangle\langle 001\mid).
\end{eqnarray}

Now we check whether entanglement is broadcasted. Making use of  Eq.(2)-(8), we can obey nonzero
 entanglement tensors  in the output state $M_{xxx}= 7/27$, $M_{xyy}=M_{yxy}=M_{yyx}=- 7/27$ and 
 $M_{zz}(1,2)=M_{zz}(1,3)=M_{zz}(2,3)= 4/9$ 
while all other entanglement tensors vanish as well as the coherence vectors.
The values $E_3$ and $E_2(m,n)$ are  then given by
\begin{equation}
 E_3=\frac{49}{729},\quad E_2(1,2)=E_2(2,3)=E_2(1,3)=\frac{16}{243}.
\end{equation}
which indicate that   both the inter-three-qubit entanglement  and inter-two-qubit entanglement in a tree-qubit systems 
charactrized by  $E_3$ and $E_2(m,m)$, respectively,   can be broadcasted via local quantum cloners, although both the 
inter-three-qubit entanglement  and inter-two-qubit entanglement   of the copied state are  less than those of 
the original state which are given by Eq.(14).


Secondly, we   consider the case  of non-local cloning.  
In this case, the entangled state of the three-qubits is treated as a state in
a larger Hilbert space and cloned as a whole. The non-local quantum copying machine  [12] 
is an $N$ dimensional quantum system,
and we shall let $\mid X_i\rangle_x$ ($i=1,\cdots,N$) be a set of orthonormal basis
of the copying machine Hilbert space. This copier is initially prepared in a
particular state $\mid X\rangle_x$. The action of the cloning transformation
can be specified by a unitary transformation acting on the basis vectors of
the tensor product space of the original quantum system $\mid \phi_i\rangle_{a_0}$, 
the copier, and an additional $N$-dimensional system which
is to become the copy (which is initially prepared in an arbitrary state
$\mid 0\rangle_{a_1}$). The corresponding transformation $U_2$ is given by
\begin{eqnarray}
  U_2\mid\phi_i\rangle_{a_0}\mid 0\rangle_{a_1} \mid X\rangle_x
 &=&c \mid \phi_i \rangle_{a_0} \mid \phi_i \rangle_{a_1} \mid X_i \rangle_x  \nonumber \\
 & &+d\sum_{j\neq i}^N(\mid\phi_i \rangle_{a_0} \mid \phi_j \rangle_{a_1} \nonumber \\
 & &+ \mid \phi_j \rangle_{a_0} \mid \phi_i \rangle_{a_1}) \mid\!X_j\rangle_x,
\end{eqnarray}
where $ i=1,\cdots,N$, $c^2\!=\!2/(N+1)$,  and $d^2\!=\!1/2(N+1)$.
 
For a three-qubit system, we have  eight basis vectors $\mid\phi_1\rangle=\mid 000\rangle$, $\mid\phi_2\rangle=\mid 001
\rangle$, $\mid\phi_3\rangle=\mid 010\rangle$, $\mid\phi_4\rangle=\mid 011
\rangle$, $\mid\phi_5\rangle=\mid 100\rangle$, $\mid\phi_6\rangle=\mid 101
\rangle$, $\mid\phi_7\rangle=\mid 110\rangle$, and  $\mid\phi_8\rangle=\mid 111
\rangle$. So that the original three-qubit GHZ state (15) can be simply  expressed as
$\mid\psi\rangle=(\mid\phi_1\rangle+\mid \phi_8\rangle)/\sqrt{2}$.
The copying is now performed by the transformation  (21)
with $N=8$. After cloning, we get the total output state:  
\begin{equation}
\mid\psi\rangle_{total}^{(out)}=\frac{1}{\sqrt{2}}U_2[(\mid \phi_1\rangle_{a_0}+
\mid \phi_8\rangle_{a_0})\mid 0\rangle_{a_1}\mid X\rangle_x].
\end{equation}

Then performing trace over the cloner, from Eq.(22) we obtain a six-qubit density operator
$\hat{\rho}_{1_01_12_02_13_03_1}^{(out)}$ which also describes two nonlocal
three-qubit systems $\hat{\rho}_{1_02_03_0}$ and $\hat{\rho}_{1_12_13_1}$.
These two three-qubit systems are the clones of the original three-qubit
GHZ state in Eq.(4.1) and they are described by the same density operator as:
\begin{eqnarray}
\hat{\rho}_{1_02_03_0}^{(out)}&=&\frac{1}{3}(\mid 111\rangle\langle 111\mid
+\mid 000\rangle\langle 000\mid)+\frac{5}{18}(\mid 111\rangle\langle 000\mid  \nonumber \\
& &+\mid 000\rangle\langle 111\mid) +\frac{1}{18}(\mid 110\rangle\langle 110\mid+\mid 011\rangle\langle 011\mid \nonumber \\
& &+\mid 101\rangle\langle 101\mid+\mid 100\rangle\langle 100\mid   +\mid 010\rangle\langle 010\mid \nonumber \\
& &+\mid 001\rangle\langle 001\mid)
\end{eqnarray}

For the output state (23) we find  the nonzero entanglement tensors $M_{xxx}=5/9$, 
$M_{xyy}=M_{yxy}=M_{yyx}=-5/9$, and $M_{zz}(1,2)=M_{zz}(1,3)=M_{zz}(2,3)= 5/9$, 
While all other entanglement tensors vanish as well as the coherence vectors.
Then the values the entanglement measures $E_3$ and $E_2(m,n)$ are are found to be
\begin{equation}
 E_3=\frac{25}{81},\quad E_2(1,2)=E_2(2,3)=E_2(1,3)=\frac{25}{243},
\end{equation}
which implies  that   entanglements are broadcasted through nonlocal quantum copying,
but  entanglement of the copied state is less than that of the original state.
Comparing Eq.(24) with Eq.(20), we   see that non-local cloning is much efficient than the local copying
process for  broadcasting entanglement.

 
Finally, it is interesting to compare the fidelity $F_1$ of the output density operator
after local copying relative to $\mid\!\psi\rangle$ with the fidelity $F_2$
of the output density operator after non-local copying relative to
$\mid\!\psi\rangle$. The fidelity of a density matrix $\rho$ relative to
$\mid\psi\rangle$ is defined by $F=\langle\psi\mid\rho\mid\psi\rangle$.
Through calculating, we find that $F_1=91/216$ and $F_2=11/18$.   Therefore,  we can see that  the output state after non-local
copying is closer to the original state than the output state after local copying.


In conclusion,  we have proposed entanglement measures for a three-particle system 
and applied  them to the study of broadcasting of entanglement for a thrr-qubit GHZ state. 
By using local and non-local cloning transformations, we have shown
that entanglement of the three-qubit GHZ state can be locally or no-locally copied
with the help of local quantum copiers or non-local quantum copiers,
and that the degree of entanglement of the resultant copies is reduced.
And we have found  that non-local cloning is much more  efficient than local cloning 
for the broadcasting of entanglement.

 \begin{center}
 {ACKNOWLEDGMENTS}
 \end{center}

This work was supported in part  NSF of China, the Excellent Young-Teacher 
Foundation of the Educational Commission of China, ECF  and STF of Hunan 
Province.

\vspace{1cm}
$^{\dagger}$Email: lmkuang@sparc2.hunnu.edu.cn

\end{document}